\documentclass[preprint]{aastex631}
\usepackage{xcolor}
\usepackage{ulem}
\bf

\shorttitle{QUASAR MICROLENSING STATISTICS}
\shortauthors{Mediavilla et al.}
\graphicspath{{./}{figures/}}

\begin{document}

\title{{Quasar Microlensing Statistics and} Flux-Ratio Anomalies in Lens Models}

\author[0000-0003-1989-6292]{E. Mediavilla}
\affiliation{Instituto de Astrof\'{\i}sica de Canarias, V\'{\i}a L\'actea S/N, La Laguna 38200, Tenerife, Spain}
\affiliation{Departamento de Astrof\'{\i}sica, Universidad de la Laguna, La Laguna 38200, Tenerife, Spain}
\author[0000-0001-7798-3453]{J. Jim\'enez-Vicente}
\affiliation{Departamento de F\'{\i}sica Te\'orica y del Cosmos, Universidad de Granada, Campus de Fuentenueva, 18071 Granada, Spain}
\affiliation{Instituto Carlos I de F\'{\i}sica Te\'orica y Computacional, Universidad de Granada, 18071 Granada, Spain}

%
%
%
%
%




\author[0000-0003-4446-7465]{V. Motta}
\affiliation{European Southern Observatory, Alonso de C\'ordova 3107, Vitacura, Santiago, Chile.}
\affiliation{Instituto de F\'{\i}sica y Astronom\'{\i}a, Facultad de Ciencias, Universidad de Valpara\'{\i}so, Avda. Gran Breta\~na 1111, 2360102 Valpara\'{\i}so, Chile}

\begin{abstract}

Precise lens modeling is a critical step in time delay studies of multiply imaged quasars, which are key for measuring some important cosmological parameters (specially $H_0$). However, lens models (in particular those semi-automatically generated) often {show discrepancies with} the observed flux-ratios between the different quasar images. These flux-ratio anomalies are usually explained through differential effects between images (mainly microlensing) that alter the intrinsic magnification ratios predicted by the models. To check this hypothesis, we collect direct measurements of microlensing 
to obtain the histogram of microlensing magnifications. We compare this histogram with recently published model flux-ratio anomalies and conclude that they cannot be statistically explained by microlensing. The average value of the model anomalies ($0.74\,$magnitudes) significantly exceeds the mean impact of microlensing ($0.33\,$magnitudes). Moreover, the histogram of model anomalies presents a significant tail with high anomalies ($|\Delta m| \ge 0.7$ magnitudes) which is completely unexpected from the statistics of microlensing observations. Microlensing simulations neither predict the high mean nor the fat tail of the histogram of model anomalies. { We perform several statistical tests which exclude that microlensing can explain the observed flux-ratio anomalies (although Kolmogorov-Smirnov, which is less sensitive to the tail of the distributions, is not always conclusive)}. {Thus, microlensing cannot statistically explain the bulk of flux-ratio anomalies, and models may explore different alternatives to try to reduce them. In particular, } we propose to complement photometric observations with accurate flux ratios of the broad emission lines obtained from integral field spectroscopy to check and, ideally, constrain lens models. 

%
  
\end{abstract}



\section{Introduction} \label{sec:intro}

Systems of multiple images of distant quasars formed by the gravitational field of intervening galaxies are one of the most useful "laboratories" in astrophysics and cosmology, allowing to study the structure of the quasar sources, the properties of matter in the lens galaxies, and the cosmological parameters, among other applications. A necessary step in these studies is the modeling of the lens, which needs to be very precise and robust, particularly for cosmographic applications like the prediction of gravitational time delays between the images, which may be used to solve the current tension in the determination of the Hubble constant, $H_0$, from different methods (see, e.g., Di Valentino et al. 2023 and references therein).
On the other hand, the number of observed systems which will need to be modeled is expected to increase considerably in the near future, which will prevent a detailed individual modeling, calling for (semi-)automated procedures (see, e.g. Shajib et al. 2019, Schmidt et al. 2023 and references therein). 

Common observable photometric/astrometric quantities of the lensed systems are the positions of the lens and images, and the fluxes of the images.
Models must, therefore, take into account the structure of the lens (often including secondary lenses), the structure of the source, and a careful modeling of the point spread function (Koopmans et al. 2003, Suyu et al. 2010, Birrer et al. 2022). To avoid an unmanageable large number of unknowns, the lens mass distribution is usually parametrized (e.g. as a power law or Navarro-Frenk-White profile). Spectroscopic information is less frequently used in spite that it may be crucial to break important degeneracies present in lens modeling.

Photometry based lens models are usually much more constrained by the astrometric observables  (because they are accurate, with typical uncertainties of a few milliarcseconds) than by fluxes. Moreover, in the case of quasar images, broad- or narrow-band fluxes can be, in principle, affected  by several sources of uncertainty like intrinsic variability of the source combined with time-delays, micro and millilensing, extinction, etc. (see Pooley et al. 2007, Yonehara et al. 2008, Motta et al. 2012 and references therein) which make them practically of no use in constraining the models. 
In fact, calculated models very often present strong differences between their predicted flux ratios and the observed ones (Witt et al. 1995, Mao \& Schneider 1998, Chiba 2002, Metcalf \& Madeau 2001, Dalal \& Kochanek 2002, Schechter \& Wambsganss 2002, Keeton 2002, Bradac et al. 2002, Metcalf \& Zhao 2002, Moustakas \& Metcalf 2003, Metcalf \& Amara 2012, Xu et al. 2009, 2015, Gilman et al. 2017). As a statistically representative case, Shajib et al. (2019) find strong flux-ratio anomalies, which they attribute mainly to microlensing, in a sample of {13} quadruple imaged quasars studied to devise a general framework to model multiply imaged quasars (with the aim of processing the large number of systems to be discovered in deep wide-field surveys like the Wide-Field Infrared Survey Telescope, LSST or Euclid). If microlensing is the cause of the anomalies, continuum flux-ratios in the visible are basically useless to model lens systems, but if the impact of microlensing cannot explain the flux-ratio departures from the predictions, {it would make sense to explore possible ways to improve the models (like the use of spectroscopic data)}.
Microlensing magnification of the quasar images can be directly (independently from lens modeling) measured  by using spectroscopic information of the quasar images. Microlensing is size sensitive (the larger the size, the smaller the impact of microlensing) and in the spectrum of each image we have information from different regions in the quasar: the continuum comes from the tiny accretion disk, which can be strongly affected by microlensing, while the broad emission lines come from the relatively large broad line region, which is rather insensitive to this effect (see, e.g,  Wisotzki et al. 1993, Mediavilla et al. 2009, 2011 and references therein). Consequently, we can use the flux ratios corresponding to the emission lines as zero microlensing baseline to measure the impact of microlensing in the continuum flux ratios at a given epoch (single epoch microlensing measurements). 

Alternatively, microlensing can also be studied from photometric monitoring of the images of a lensed quasar. Subtracting the (time delay corrected) light curves of two images, we can obtain microlensing light curves. Owing to the presence of extinction and to the relatively reduced extension of the monitoring period we can not fully quantify the total amplitude of microlensing at a given epoch, but the amplitudes of the peaks of microlensing events should provide a conservative upper bound to single epoch microlensing.

The first objective of this paper is, therefore, to obtain the histogram of observed flux-ratio anomalies induced by microlensing. In order to do so, we calculate the experimental differential microlensing magnifications for a sample of  {44 measurements in 34  image pairs of 23 lens systems} with available spectroscopic information (collected by Esteban-Guti\'errez et al. 2022 {from Rojas et al. 2020, 2014, Motta et al. 2017, 2012, Jiménez-Vicente et al., 2015, and Mediavilla et al. 2009}).   We then compare this experimental histogram with the flux-ratio anomalies inferred from model predictions in two samples of quadruple lens systems (Shajib et al. 2019, Schmidt et al. 2023),  to {illustrate how} the hypothesis that the flux-ratio anomalies are caused by microlensing {can be tested. The present work does not intend to make any general statement about lens modeling. Instead, we just aim at providing some tools to detect potential problems in some models and to, eventually, improve them in some cases.}

%
%
%

The paper is organized as follows. In \S2 we present the histogram of observed microlensing magnifications (based on spectroscopic data), and compare it with the statistics of microlensing peak amplitudes (derived from microlensing light curves) and with the predictions of microlensing simulations. In \S3 we collect histograms of microlensing model anomalies from the samples in Shajib et al. (2019) and Schmidt et al. (2023) and compare them with the histograms of observed microlensing magnifications and of microlensing peak amplitudes. In \S4 we discuss possible observational strategies to derive useful constraints based on the flux ratios either to cross-check the models or to improve them. In \S5 we summarize the main conclusions. Finally, we devote an Appendix to explore the relationship of flux-ratio anomalies with the degeneracy of lens models with respect to the radial mass distribution.

\section{Observed microlensing in multiple imaged quasars} \label{sec:obs}
\subsection{Estimates from emission-lines}
\begin{figure}[ht!]
\plotone{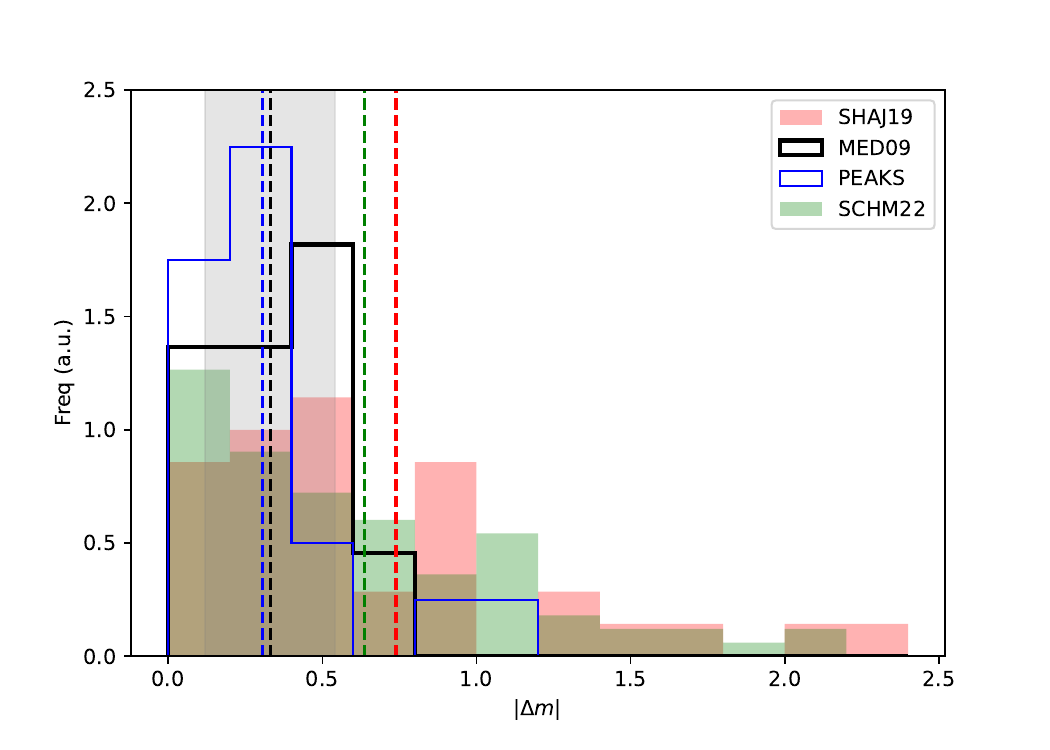}
\caption{Flux-ratio histograms corresponding to: single epoch microlensing obtained using the broad emission lines as reference (thick, black line), peaks in microlensing light curves (thin, blue line), models from Shajib et al. (2019) (red shaded histogram), and models from Schmidt et al. (2023) (green shaded histogram). Vertical lines show the mean values. The grey shaded region corresponds to one standard deviation around the mean for single epoch microlensing and microlensing peaks.
  \label{figure1}}
\end{figure}

To directly measure the impact of microlensing in the images of lensed quasars, we can take advantage of the sensitivity of microlensing to the size of the source. Microlensing by a distribution of stars induces strong spatial changes  ("microlensing roughness") in the otherwise uniform (smooth) magnification at the source plane. If the size of the source is large enough,  the inhomogeneities of the magnification are spatially averaged and washed out. The spatial scale of the magnification roughness is related to the Einstein radius  of the microlenses, which for a  typical mass of 0.3$M_\odot$ and typical values for the redshifts of the lens ($z_l=0.5$) and the source ($z_l=2$)  amounts  to approximately $10\,$light-days. Consequently, the impact of microlensing can be potentially high for the quasar continuum source  (a few light-days in size) but negligible for the Broad Line Region (BLR) (with sizes above hundred light-days) (e.g., Jim\'enez-Vicente et al. 2022, Guerras et al. 2013, Fian et al. 2018). Thus, we can use the broad lines present in quasar spectra (in particular, the core of the lines) to determine the zero microlensing baseline. For a pair of images of a lens system, we can define the relative microlensing magnification between them as the continuum ratio relative to the zero point defined by the emission line ratio. Expressed in magnitudes we can write (see, e.g., Mediavilla et al. 2009),

\begin{equation}
\label{micro}
\Delta m_{ij}=(m_i-m_j)^{cont}-(m_i-m_j)^{line.},
\end{equation}
Using  the sample of 44 microlensing measurements {collected by} Esteban-Guti\'errez et al.\footnote{Collected from Rojas et al. 2020, 2014, Motta et al. 2017, 2012, Jiménez-Vicente et al., 2015, and Mediavilla et al. 2009} (2022) obtained according to Eq. \ref{micro}, we derive the histogram of (unsigned) microlensing magnifications (shown in Figure \ref{figure1}).  The average rest wavelength for which these measured differential microlensing magnifications are estimated is $\lambda \sim 1700$ \AA\, (cf. Jim\'enez-Vicente et al., 2012). The histogram has a mean 
of $\langle |\Delta m_{ij}|\rangle=0.33\pm 0.22$. This is, in fact, an overestimate of the expected impact of microlensing, as all four single measurements in Eq.\ref{micro} are also affected by experimental uncertainties, which will broaden the intrinsic histogram of microlensing magnifications. {We have estimated directly the mean error, $\langle\sigma_{\Delta m}\rangle$ from the different data sources used: $0.13\pm 0.09$ (Mediavilla et al.\footnote{There is no rms estimates for all the measurements.} 2009), $0.11\pm 0.04$ (Motta et al. 2011), $0.23\pm 0.02$ (Rojas et al. 2014), $0.15\pm 0.13$ (Motta et al. 2017), and $0.11\pm 0.07$ (Rojas et al. 2020). The weighted average of the errors is $0.13$. }

%

Notice that the experimental $\Delta m_{ij}$ also includes the differences in flux arising from intrinsic quasar variability combined with the time delay between images, which are supposed to be small, specially for quads. On the contrary, as far as the continuum and emission lines are observed at close wavelengths, we can assume that the $\Delta m_{ij}$, calculated using Eq.\ref{micro}, are virtually free from extinction. 

{In the histogram of microlensing magnifications (Figure \ref{figure1}) doubles and quads are mixed. If we separate both groups, we obtain $\langle |\Delta m_{ij}|\rangle_{quads}=0.27\pm 0.22$ and $\langle |\Delta m_{ij}|\rangle_{doubles}=0.40\pm 0.19$. The results indicate that doubles exhibit slightly larger microlensing than quads (likely because doubles have larger time delays, which combined with intrinsic variability may increase flux-ratio anomalies; on the other hand in doubles one of the images is also often close to the lens galaxy and, consequently, more prone to microlensing). Then, if we restrict to quads, micro-lensing anomalies would be even slightly smaller. Some of the microlensing measurements in our sample correspond to different epochs of a same image pair. To avoid the impact of possible covariance between repeated measurements we have substituted for each image pair where more than one measurement were available, all the measurements by its mean, finding neglectable differences in $\langle |\Delta m_{ij}|\rangle$ either for quads or doubles.}

\subsection{Comparison with peak amplitudes of microlensing light-curves}
It is interesting to compare the values of microlensing magnifications obtained using the emission lines as baseline, with the peak amplitudes of microlensing light curves. In Figure   \ref{figure1}, we also present the histogram of microlensing peak amplitudes taken from Mediavilla et al. (2016). These amplitudes use the "flat" part of the light-curve before or after the microlensing event as baseline. This is an idealization, because due to the relatively high optical depth, quasar light curves can not generally be described as isolated microlensing events/peaks over a flat baseline. In fact,  it is common that microlensing light-curves do not present a well defined flat region. Moreover, the peak or part of the baseline can fall in one of the (seasonal or incidental) gaps of the light-curves. For this reason, the peak amplitude defined with respect to the left or right sides from the peak, can be different in some cases. We have selected always the largest one. 

There is a great similarity between the histograms of peak amplitudes and of microlensing magnifications from emission lines, although an offset towards larger values of the histogram of microlensing peak amplitudes would be expected. 
The coincidence of the means of both histograms can result, in part, from the above mentioned overestimate in the mean value of microlensing inferred from the emission lines due to measurement uncertainties. Notice also that intrinsic variability is contributing to the microlensing magnifications from emission lines while it is not affecting to microlensing light curves which are obtained subtracting time delay corrected  light curves of two images. Moreover, an underestimate in peak amplitudes can be produced because the true microlensing zero-point may fall below the "flat" regions of the microlensing light-curves taken as baseline\footnote{In not long enough light curves, caustic clustering can mimic a fake "flat" region.}, or because the maximum is in a seasonal gap.

Finally, note that for source sizes comparable to the Einstein radius of the microlenses, lensed quasar images can be frequently engaged in microlensing events with gentle slopes and broad peaks of relatively small amplitude. In fact, taking into account the total monitoring time (310 years) of the light curves of the ensemble from Mediavilla et al. (2016), the total number of microlensing events detected in the ensemble (20) and the mean Einstein radius crossing time scale\footnote{Calculated using Eq.  8 of Mediavilla et al. 2016 for a typical lens system: $z_{lens}=0.5$, $z_{source}=2$, $\sigma_*=200\rm\, km\,s ^{-1}$, and $\sigma_{pec}(z=0.5)=638\rm\, km\,s ^{-1}$. } of $9.4$ years, a 61\% of the images will be engaged in a microlensing event at any time (slightly above the  50\% estimate by Mosquera \& Kochanek, 2011). Thus, many of the measurements from emission lines likely correspond to images undergoing a microlensing event with amplitudes not very different from that of the peak. Although, even with all these considerations in mind, the coincidence between the means obtained either from the line-emission or from the light-curves may remain questionable, the absence of a significant high magnification tail in both histograms is a very robust common result.


\subsection{Theoretical microlensing estimates using reverberation mapping sizes for the quasar source}

It is also possible to make a theoretical estimate of the expected impact of microlensing from simulations based on microlensing magnification maps. The key parameter in the simulations is the size of the continuum quasar source, which can  be estimated around $r_s= 5\rm\, light-days$ from reverberation mapping studies (see, e.g., Edelson et al. 2015, Fausnaugh et al. 2016, Jiang et al. 2017, Esteban Guti\'errez et al. 2022 and references therein). Taking this value for $r_s$, Esteban-Guti\'errez et al. (2022) calculate the probability distributions of microlensing magnifications corresponding to a population of stars, for all the objects in the sample used in the present work. As it can be observed (see the blue lines in their Figure A1), the impact of microlensing is concentrated around zero, with typically, $\sigma(\Delta m) \le 0.4$,  and with a negligible tail above $|\Delta m|>1$, in agreement with the emission-line based measurements.

\section{Comparison with model flux ratio anomalies} \label{sec:floats}


 {  In order to compare the observed microlensing flux ratios with model predictions, we consider here the work of Shajib et al. 2019 (and its extension by Schmidt et al. 2023, see below). These authors explicitly introduce the question of flux-ratio anomalies related to microlensing and provide a homogeneous data set which has been modeled in a very systematic way. On the other hand, this work has the interesting perspective of exploring semi-automated modeling to face the future massive data availability.} In Figure \ref{figure1} we include the histogram of frequencies of (unsigned) flux-ratio anomalies obtained from Shajib et al. (2019) corresponding to a sample of  {13} quads.
{We use this sample just as a test bench to illustrate how to check the impact of microlensing knowing that no general conclusion about lens modeling can be inferred from a particular set of models that the authors themselves consider susceptible of refinement in several ways. We hope that the present work can be one of them.} 

{In each quad we take image A as reference to compute the magnitude differences}. We have used the data in the F475X filter from Shajib et al. (2019), which have the closest wavelength correspondence with the average rest wavelength of the microlensing measurements described in Section \ref{sec:obs}\footnote{ The average rest wavelength (see Jiménez-Vicente et al. 2012) of the used emission lines is $1736\pm373$\AA, which for a mean redshift $z=1.9$ results in an observed wavelength of $5034.4$\AA, close to the central wavelength, $4940.7$\AA, of the 475X HST filter.}. The mean of the histogram, $\langle|\Delta m_{models\_shajib}|\rangle=0.74$, greatly exceeds the mean of the microlensing measurements from emission lines ($\langle|\Delta m_{lines}|\rangle=0.33$). On the other hand, comparison of the tails of the histograms, also reveals strong differences. The high magnification tail is very populated in the case of the model flux anomalies (36\% of pairs with  $|\Delta m_{models\_shajib}| \ge 0.74$ magnitudes) while there is only one case  (2.2\%) in the histogram of microlensing from emission lines, and two (10\%) in the sample of microlensing peaks. Finally, we have computed several statistical tests that reject the hypothesis that this sample of flux ratio anomalies and the sample of direct microlensing measurements estimated from the emission lines have the same parent population {(see Table \ref{tab:tests})}. { The same negative conclusion is reached using the sample of microlensing estimates from the light curves peaks (Table \ref{tab:tests}).}

Very recently, the sample by Shajib et al. (2019) has been enlarged with 16 additional systems by Schmidt et al. (2023), who compute new models for all the quads.  In Figure \ref{figure1} we include the histogram of anomalies corresponding to Schmidt et al. (2023), which  presents an offset mean  $\langle |\Delta m_{models\_schmidt}|\rangle=0.64$ and a populated high magnification tail (30\% of pairs with $|\Delta m_{models\_schmidt}| \ge 0.74$ magnitudes). Although the results confirm those obtained from Shajib et al. (2019) sample, { the differences are slightly smaller}. { The performed statistical tests (see Table 1) reject the hypothesis that model flux ratio anomalies and observed microlensing anomalies come from the same underlying distribution. Only the Kolmogorov-Smirnov test (known to be less sensitive than the others, particularly to differences in the tails of the distributions) is inconclusive for the case of the comparison of the observed sample based on emission lines with the results by Schmidt et al. (2023).}

%


{It is interesting to notice that the average redshifts of the lens galaxies (present sample: $0.53\pm 0.18$; Shajib et al sample: $0.44\pm 0.14$;  Schmidt et al. sample: $0.47\pm 0.23$) and of the quasar sources (present sample: $1.90\pm 0.54$; Shajib et al sample: $2.10\pm 0.75$;  Schmidt et al. sample: $2.20\pm 0.82$), are statistically consistent.}

{Finally, notice that early calculations of flux-ratio anomalies by Pooley et al. (2007) with a smaller (10 systems) sample result in $\langle|\Delta m_{models\_pooley}|\rangle=0.59$ with a 24\% of pairs with  $|\Delta m_{models\_pooley}| \ge 0.74$, consistent with the results corresponding to Shajib et al. (2019) or even larger taking into account that Pooley et al. photometric data correspond to $\sim 8000\rm\, \AA$ (observed).}

Therefore, from the statistical comparison, we can conclude that the flux ratio anomalies inferred from the lens models that we have taken as example, can not be attributed to microlensing, and that{, consequently, there is room to improve the flux-ratio predictions of the models and reduce the bulk of the anomalies.}


\section{Discussion} \label{sec:discussion}

\subsection{Statistical cross-check}

According to the previous analysis, the {flux-ratio predictions of the} considered lens models do not pass the statistical cross-check based on quasar images {observations}. To estimate the true amplitude of the deviations of the model predictions, we can remove\footnote{Considering that the involved parent PDFs are one sided-normal distributions.} the expected average effect of microlensing, after which we are left with a mean model-predicted flux-ratio anomaly of 0.66 magnitudes for Shajib et al. (2019) models (0.55 for Schmidt et al. 2023 models) which can neither be attributed to microlensing nor to intrinsic source variability\footnote{Because, as commented before, the random effects of quasar variability combined with time delay are also included in the microlensing estimates based on Eq.\ref{micro}.}.

A mean deviation of 0.66 magnitudes is so large, that other causes frequently invoked to explain the flux-ratio anomalies can be confidently ruled out according to previous estimates from the literature (see Motta et al. 2012, Pooley et al. 2007). To confirm this directly from the data, we repeat the histograms removing 12 image-pairs where extinction can play a significant role finding that the high magnification tail remains. On the other hand, to check the possible impact of time-delay in the flux-ratio anomalies we have compared the histograms considering all the image-pairs in the samples with histograms excluding pairs with time delays greater than 10 or 30 days, respectively, finding no significant differences among them. In fact, following the steps described in Yonehara (2008) and assuming a time delay of 30 days and an absolute demagnified magnitude of the sources in the -21 to -23 magnitude range, we find an uncertainty of just 0.26 magnitudes in the worst case.

{Millilensing (by dark matter subhaloes, for instance) may also, in principle,  contribute to flux-ratio anomalies. However, Pooley et al. (2007, 2012) find that X-ray anomalies are much larger than the optical ones. This result, confirmed by Jimenez-Vicente et al. (2015), indicates that the effect causing the anomalies is sensitive to the differences in size between the X-ray and optical sources, while they should perform as point like under the lensing action of  large mass millilenses (Pooley et al. 2007, 2012). Based on a similar reasoning, Pooley et al. (2007) also exclude  that changes in the smooth lens model component can explain at once the anomalies in X-ray and in the optical.}

Then, the large anomalies are, indeed, an indication that  {there is room to improve} {flux predictions of} lens models. 
In particular, an examination of the specific procedure followed to fit the model for each system should be performed to analyze the origin of the discrepancies in the fluxes and their relationship with possible uncertainties in the time delay estimates for cosmographic studies. {It is possible that with a small effort of sophistication in the models, the predictions of the flux-ratios improve drastically (Ertl et al. 2023), though, the impact of these changes on the time-delays should be, anyway, examined.} {Although lens modeling of specific systems is outside of the scope of this work, in Appendix \ref{model} we explore flux-ratio anomalies under some simplifying assumptions which, while they may not reproduce the complexity of the real problem, may  still provide some interesting insight.}

\subsection{Individual model cross-check and modeling using (integral field) spectroscopic flux-ratios}

Perhaps the most interesting reflection to improve lens modeling is that accurate experimental determinations of intrinsic flux ratios (free from the effects of microlensing and extinction) obtained from the broad emission lines of quasars can be used to cross-check individual models\footnote{Incidentally,  extinction can be disentangled from microlensing by using emission lines at different wavelengths.}. Moreover, for quads, the effects of variability combined with time delays between images can be reasonably controlled (in particular when a determination of the time delays is available). In principle, narrow emission lines, mid-infrared or radio emission may also be used to determine the intrinsic flux ratios free from microlensing, but as far as the emitting regions involved are much larger than the continuum source, their images might present different flux ratios depending on the shape, centroid, extension and location of the source respect to the macro caustic. 
Notice, also, that only the use of emission lines of wavelengths close to the continuum as baseline automatically cancels the effects of extinction (see Eq.\ref{micro}). Accurate flux-ratios free from these systematic effects can be used to select reliable systems for cosmographic studies, rejecting those systems with large anomalies.

The emission lines of lensed quasars are relatively bright, and Integral Field Spectroscopy (IFS) combined with adaptive optics in large telescopes can be a very reasonable experimental possibility to simultaneously observe the emission in the continuum and in the lines of a  large enough ($\sim50$) sample of lensed systems as to estimate $H_0$ with a few percent precision\footnote{Assuming that systematic effects are not biasing the estimates.}. For instance,  a 20 magnitude object  (I filter) can be observed with HARMONI@ELT (Thatte et al. 2016, 2020), with  S/N $\sim 100$ with a total time exposure $\lesssim 10\rm\,min$, with an spatial resolution of $\sim10\,\rm mas$ and a spectral resolution of 0.208$\,\rm nm$ (adjacent wavelength slices could be co-added to further increase the S/N ratio). Thanks to the  broad spectral range covered with HARMONI, several emission lines and continuum bands can be observed at once, and it is possible to perform a simultaneous fitting of the lens system images in all of them, significantly increasing the reliability and robustness of the analysis.

Going a step further, can the flux ratios inferred from the emission lines be used not only to cross-check individual models but also to effectively constrain them? Commonly adopted uncertainties for continuum based flux-ratios between images (of even 20\%) make their use irrelevant as compared with that of the astrometry and, likely for these reason, the use of flux-ratios has been, in general, considered accessory. However, with IFS based flux-ratios with a few percents of relative uncertainties, the role of flux-ratios may be interesting to constrain the lens models\footnote{Of course, other type of ancillary observations (e.g. data from stellar kinematics, Kochanek, 2020) can provide the information needed to solve the degeneracies.}.

{In the simple study performed by us to explore the impact of the radial dependence of the gravitational potential (see Appendix \ref{model}), we see that, in most cases, flux ratios are much less sensitive than time delays to changes in the potential, although the impact in some image-pairs of some specific systems may be large enough as to help breaking the degeneracy.
In any event, we have used a very simple model which, among other issues, does not take into account any complexity of the angular part. A more thorough exploration of lens modeling is needed to ascertain the real usefulness of precise flux ratios to constrain the models and improve the robustness of theoretical time delay estimates.}

\section{Summary and conclusions} \label{sec:extinction}

We use a sample of  {44 measurements from 34 image pairs of 23} lensed systems with spectroscopic observations to obtain  the histogram of  microlensing magnifications, using the emission lines to define the non-microlensed baseline. This histogram can be used to perform a statistical cross-check of lens models comparing with the deviations of the predicted flux-ratios with respect to the observed ones. To illustrate this possibility we obtain the histogram of model flux-ratio anomalies (predicted minus observed flux-ratios) from Shajib et al. (2019) and Schmidt et al. (2023). The main conclusions are the following:

1 - The mean value of the model anomalies ($\langle|\Delta m_{models}|\rangle= 0.74$) exceeds significantly the mean impact of microlensing ($\langle|\Delta m_{lines}|\rangle= 0.33$). The histogram of model anomalies shows a significant tail ($|\Delta m_{models}|\ge$ 0.7 magnitudes) not present in the histogram of directly measured microlensing magnifications. 
The histogram of peak amplitudes of microlensing events obtained from microlensing light curves, neither presents this extended tail. These results strongly disfavors the hypothesis that the model flux-ratio anomalies arise mainly from microlensing. 

2 - Consequently, the remaining flux-ratio anomalies (after removing microlensing and intrinsic {variability combined with time delay} effects) of $\langle |\Delta m |\rangle =0.6$ magnitudes {may be reduced by further refinements of the models, which are} well outside of the scope of the present work. {Using, nevertheless, an exploratory simple model, we find that the degeneracy of astrometric model fitting with the radial distribution of mass in the lens can account only for a relatively small part of the observed flux-ratio anomalies (departures from ellipticity of lens galaxies can play a more significant role).}


3 - In principle, models can be cross-checked, not only statistically but also individually, using flux ratios from emission (mid infrared, radio, broad and narrow emission lines, for instance) coming from regions large enough as to be insensitive to microlensing. However, if the region is too large (including the dusty torus, the NLR or the radio jet, for instance), additional modeling of the source and of its (extended) images is needed. In this sense, the use of the relatively compact BLR can be less complex. We propose to use spectroscopic data, specifically based on integral field spectroscopy, to measure with current (SINFONI, MUSE) or future (HARMONI) instrumentation, accurate broad emission line fluxes, to obtain flux-ratios free from microlensing and extinction to check the models just to the experimental uncertainties of flux photometry. 

4 - In most cases, the uncertainties associated to the possible impact of microlensing in the observed continuum flux ratios, have made them irrelevant in lens model fitting. The use of very accurate broad emission line flux-ratios to establish effective constrains in the models should be explored.

Finally, we recommend the consideration of the flux-ratio anomalies as a quality check for the fitted models, and we advice to discard those systems with unexpectedly large anomalies. Although a large flux-ratio anomaly in an individual system is certainly not warranty of a wrong model, it may be a good warning signal which is worth taking into account. Given that ongoing and future surveys will produce large numbers of lensed systems suitable of being used for cosmographic studies, discarding a fraction of suspicious systems shall not damage the statistical quality of those studies. 

\begin{acknowledgments}
This research was supported by grants PID2020-118687GBC33 and PID2020-118687GB-C31, financed by MCIN/AEI/10.13039/501100011033. J.J.V. is also financed by projects FQM-108, P20\_00334, and A-FQM-510-UGR20/FEDER, financed by Junta de Andaluc\'{\i}a. V.M. acknowledges support from ANID Fondecyt Regular \#1231418 and Centro de Astrof\'{\i}sica de Valpara\'{\i}so.
\end{acknowledgments}

\appendix

\begin{figure}[ht!]
\plotone{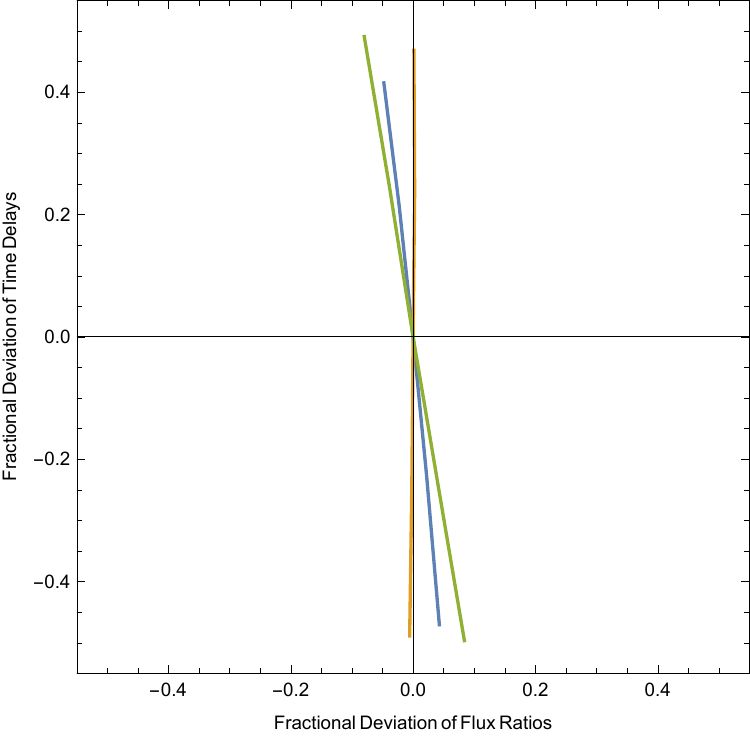}
\caption{Fractional variations of time delays vs. flux ratios from models obtained for SDSS J2048+1913 changing the exponent of the power law mass distribution of the lens in the $\beta=0.5$ to $1.5$ range. Each color corresponds to an image pair. \label{taovsflux}}
\end{figure}

{

\section{What can be learnt from model flux anomalies? \label{model}}

%

The degeneracy of models based in astrometry with respect to the law describing the radial mass distribution in the lens galaxy is an often invoked difficulty of lens models to provide accurate estimates of the true time delays (see Kochanek, 2020, and references therein). We can explore here whether the flux-ratio anomalies can also be related to this degeneracy\footnote{Because the non geometrical part of the time delays is proportional to the potential strength, they are strongly affected by the degeneration in the potential, but the question is not immediate in the case of flux-ratios, as they depend on the second derivatives of the potential.},  estimating and comparing its impact in both quantities. With this exploratory aim, we can consider the singular isothermal ellipsoid potential (SIE) generalized to take into account a power law dependence with size, $\phi_{SIE} = b\left[(x^1\cos{\theta_\epsilon}-x^2\sin{\theta_\epsilon})^2/q^2+(x^1\sin{\theta_\epsilon}+x^2\cos{\theta_\epsilon})^2 q^2\right]^{\beta/2}$, where $q^2$ is the axial ratio. Expanding to first order this potential and adding an external shear, we can write,


\begin{equation}
\phi=b|\vec r|^\beta-b{|\vec r|^2 \over |\vec r|^{2-\beta}}\gamma_{\epsilon}\cos{2(\theta-\theta_{\epsilon})}-{1\over 2} b |\vec r|^2 \gamma_{ext}\ cos{2(\theta-\theta_{ext}}),
\end{equation}
where the second term is analogous to the SIE quadrupole (see, e.g., Kochanek 2002). Using this potential, we fit the images and lens positions of the nine quadruple lens systems that Shajib et al. (2019) modeled with a single lens mass profile. We consider logarithmic slopes of the power-law in the range $\beta=0.5$ to 1.5. Eight of the nine systems are very well fitted by our simple model for all the values of $\beta$, ($\chi^2(\beta)\le 1$), confirming the degeneracy of models based on astrometric data with respect to plausible radial dependences\footnote{In the case of PS J0147+4360, a good fit of the images positions can also be achieved, but the position of the lens galaxy needs to be offset by more than 40 miliarcsec.}. Then, we compute the variation of flux ratios and time delays between images in the considered range of $\beta$. In Figure \ref{taovsflux} we illustrate, for one of the lens systems (SDSS J0248+1913), the fractional deviation of both magnitudes (time delays vs. flux ratios) with respect to a fiducial model that we arbitrarily select for $\beta=1$, corresponding to SIS+$\gamma_{\epsilon}$+$\gamma_{ext}$.  As it is shown in this Figure, the maximum deviation of the flux ratios  ranges from 0\% to $\pm$25\%  (depending on the pair of images), while a  larger maximum fractional deviation of $\pm$50\% is obtained for the time delays for all the three image pairs.  Similar results (maximum fractional variations of the flux ratios between 0 and $\pm30\rm\%$ with a mean value of about $10\%$ while the time-delays exhibit a much larger typical variation of about $50\rm\%$) are derived considering all the image pairs of the lens systems in the sample when $\beta$ is changed.

In principle, these results show that the degeneracy in the radial mass distribution of the lens may be a common source of uncertainties for flux ratios and time-delays.
However, the range of variability of flux ratios with $\beta$  would only account for a small part of the measured model anomalies ($\langle \Delta m \rangle = 0.6$ magnitudes is equivalent to a fractional deviation of 60\%). This indicates that other ingredients of the model (aside from the radial dependence of the lens potential) might be affecting the flux ratios\footnote{We are assuming that, on average, observed flux ratios are reasonably reliable, and not contaminated by extended emission or other potential sources of uncertainty.}. In fact, on top of the degeneracies on the radial dependence of the mass distribution, the angular dependence may also induce biasses in the models (e.g. Kochanek, 2021, Van de Vyvere et al. 2022, Gomer et al. 2022). Several recent works have indeed shown that assuming elliptical models can bias the estimate of $H_0$ up to 10\% (see e.g. Gomer \& Williams 2021; Cao et al. 2022) and flux-ratio anomalies have proven to be indicative of non-elliptical components in the mass distribution in some systems (e.g. Hsueh et al. 2016, 2017). The anomalies virtually disappeared when the non-ellipticity is included in the models. Then, more complex models (and ancillary data) are needed to account for the observed anomalies in the flux-ratios and to explore possible correlations with uncertainties on the time-delays.
}

\begin{deluxetable*}{lcccc}
\tablecaption{p-values for several statistical tests\label{tab:tests}}
\tablewidth{0pt}
\tablehead{
\colhead{} & \colhead{SHAJ19$^1$} & \colhead{SHAJ19}  & \colhead{SCHM23$^2$} & \colhead{ SCHM23} \\
\colhead{} & \colhead{vs. PEAKS$^3$} & \colhead{vs. MED09$^4$} 
 & \colhead{vs. PEAKS} & \colhead{vs. MED09}
}
\startdata
Kolmogorov-Smirnov & 0.01 &  0.03& 0.00 & 0.18 \\
Epps-Singleton & 0.00 &  0.00 & 0.00 & 0.00  \\
Anderson & 0.02 &  0.02 & 0.02 & 0.02  \\
Cramer-von Mises& 0.03 & 0.04  & 0.01 & 0.06  \\
\enddata
\tablecomments{$^1$Sample of flux-ratio anomalies from Shajib et al. (2019). $^2$Sample of flux-ratio anomalies from Schmidt et al. (2023). $^3$Sample of microlensing estimates based on light-curves peaks. $^4$Sample of single-epoch microlensing measurements based on emission lines. See Figure \ref{figure1}.}
\end{deluxetable*}

\clearpage

\end{document}